\documentclass[12pt]{article}
\usepackage[utf8]{inputenc}
\usepackage[english]{babel}
\usepackage[pdftex]{graphicx}
\usepackage{cite}
\usepackage{cmap}
\usepackage{authblk}
\usepackage{amssymb}
\usepackage{amsmath,mathtools}
\usepackage{amsfonts}
\usepackage[compat=1.1.0]{tikz-feynman}
\tikzfeynmanset{/tikzfeynman/warn luatex=false}
\def\dd{{\mathrm d}}
\newcommand{\SANC}{\texttt{SANC}{}}
\newcommand{\MCSANC}{\texttt{MCSANC}{ }}

\newcommand{\ReneSANCe}{\texttt{ReneSANCe}{ }}

\newcommand{\sss}[1]{\scriptscriptstyle{#1}}

\def\MSbar{\overline{\tiny \mathrm{MS}}}

\begin{document}

\title{\Large
Electroweak Effects in Neutral Current Drell-Yan Processes 
within \SANC ~System
}

\author[1]{A.\,Arbuzov}
\author[1]{S.\,Bondarenko}
\author[2,3]{Ya.\,Dydyshka}
\author[2]{L.\,Kalinovskaya}
\author[2]{R.\,Sadykov}
\author[2,3]{V.\,Yermolchyk}
\author[2,3]{Yu.\,Yermolchyk}

\affil[1]{\small Bogoliubov Laboratory of Theoretical Physics, JINR, 141980 Dubna, Moscow region, Russia}
\affil[2]{\small Dzhelepov Laboratory of Nuclear Problems, JINR, 141980 Dubna, Moscow region, Russia}
\affil[3]{\small Institute for Nuclear Problems, Belarusian State University, Minsk, 220006  Belarus}

\date{\vspace{-5ex}}
\maketitle

\begin{abstract}
The complete one-loop electroweak radiative corrections 
to the neutral current Drell-Yan  process $pp \to \ell^+\ell^- X$ 
are presented for the case of longitudinal polarization of initial
particles. The calculations are based on the {\SANC} computer system. 
The paper contains a brief description of the {\SANC} approach and 
a discussion of the sources of theoretical uncertainties in electroweak effects.
\end{abstract}

\section{Introduction}
Theoretical calculations of unpolarized Drell-Yan (DY) processes for high-energy hadronic colliders were performed 
at the level of one-loop QED and electroweak (EW) radiative corrections (RC) by several groups, see
~\cite{Mosolov:1981xk,
Soroko:1990ug,
Wackeroth:1996hz,
Baur:1998kt,
Dittmaier:2001ay,
Baur:2001ze,
Baur:2002fn,
Baur:2004ig,
CarloniCalame:2006zq,
CarloniCalame:2005vc}
and references therein. The aim of this paper is to present our new results on 
electroweak corrections to polarized DY processes within the \SANC ~computer system.

The current versions of the Monte Carlo generator {\ReneSANCe}~\cite{Bondarenko:2022mbi} and Monte Carlo integrator {\MCSANC}~\cite{Bondarenko:2013nu,Arbuzov:2015yja} are adjusted 
to studies of various effects due to EW radiative corrections in realistic LHC observables.
The series of our works 
\cite{Arbuzov:2007kp,
Bardin:2012jk,Bondarenko:2013nu,
Arbuzov:2015yja,
Antropov:2017bed,Bondarenko:2022zse}
was dedicated to the implementation of the EW neutral current (NC) DY branch into 
our Monte Carlo tools {\ReneSANCe} and {\MCSANC} based on the {\SANC} ~standard
modules~\cite{Andonov:2008ga}. 
In this note, we do not describe the {\SANC}~system itself, referring the reader to~\cite{Andonov:2004hi,Bardin:2019zsp}.

We participated in workshops on tuned comparison between programs for NC DY \cite{TeV4LHC-Top:2007fwh,Alioli:2016fum}.
The results of our calculations of NC DY were compared with the results
of other theoretical groups {\tt HORACE}~\cite{CarloniCalame:2006zq}, 
{\tt ZGRAD2}~\cite{Baur:2001ze}, 
{\tt POWHEG BOX}~\cite{Alioli:2010xd}, {\tt PHOTOS}~\cite{Golonka:2005pn} and shown excellent agreement.
The review \cite{Alioli:2016fum} was prepared as a part of the LPCC
EW Precision Measurements in WG LHC. It summarizes the work of the Systematic
Comparison of MC codes describing DY processes at the LHC.
This work was an important step toward determining the exact limits
of the Standart Model needed for high-precision measurements of EW observables
such as the W-boson mass and the weak mixing angle.

\section{Different sources of theoretical uncertainties}
The resulting theoretical uncertainty in the description of the EW part of the NC DY process should combine errors coming from several sources. 
Let us briefly review different sources of theoretical uncertainties in this process 
and their state of art in {\SANC}  MC codes.

\underline{\bf Complete one-loop  RC}

The adopted form of presentation of a differential cross-section at the one-loop level is
$\hat{\sigma}^{\mathrm{1-loop}} = \hat{\sigma}^{\mathrm{Born}} + \hat{\sigma}^{\mathrm{virt}}(\lambda)
+ \hat{\sigma}^{\mathrm{soft}}(\lambda, \omega)
+ \hat{\sigma}^{\mathrm{hard}}(\omega) + \hat{\sigma}^{\mathrm{Subt}}, 
\label{loopxsec}$
where  $\hat{\sigma}^{\mathrm{Born}}$ is due to the contribution of the Born level cross section,
$\hat{\sigma}^{\mathrm{virt}}$ is due to the virtual (loop) corrections,
$\hat{\sigma}^{\mathrm{soft}}$ is due to soft photon emission,
and  $\hat{\sigma}^{\mathrm{hard}}$ is due to hard photon emission (with energy $E_{\gamma} > \omega$).
Here we have the auxiliary parameters $\omega$ which separates soft and hard photons, and $\lambda$ --- a fictitious "photon mass" which regularizes infrared divergences. In our calculations, we always 
demonstrate the numerical independence of these auxiliary parameters.
The special term  $\hat{\sigma}^{\mathrm{Subt}}$ stands for subtraction of collinear quark mass singularities.
The calculations were performed in unitary and $R_\xi$ gauges for a cross-check.

It is also possible to evaluate some contributions separately:
QED and weak-interaction parts; various contributions to the virtual part 
(sum of self-energy and vertices, all topologies of boxes),
the initial state radiation, the final state radiation and their interference.

The differential cross section of the DY  process at the hadronic level 
can be obtained from the convolution of the partonic cross section
with the quark density functions:
\begin{eqnarray}
\label{sigpp}
&& \frac{\dd\sigma_{1}^{pp\to l\bar{l} X}(s,c)}{\dd c} 
= \sum\limits_{q_1q_2}\int\limits_{0}^{1} \int\limits_{0}^{1} 
\dd x_1\; \dd x_2\; \bar{q}_1(x_1,M^2) 
\times
\bar{q}_2(x_2,M^2)
\frac{\dd\hat{\sigma}_1^{q_1\bar{q}_2\to l\bar{l}}(\hat{s},\hat{c})}
{\dd\hat{c}}
\nonumber \\ && \qquad \times 
{\mathcal J}\Theta(c,x_1,x_2),
\end{eqnarray}
where the step function $\Theta(c,x_1,x_2)$ defines the phase space domain 
corresponding to the given event selection procedure and ${\mathcal J}$ is the relevant Jacobian.
The partonic cross section is taken in the center-of-mass reference 
frame of the initial quarks, where the cosine of the muon scattering angle $\hat{c}$ is defined.
The parton densities with {\em bars} in Eq.~(\ref{sigpp}) mean the ones
modified by the subtraction of the quark mass singularities:
\begin{eqnarray}
&& \bar{q}(x,M^2) = q(x,M^2) - \frac{\alpha}{2\pi} \, Q_q^2
\int_x^1 \frac{\dd z}{z} \, q\biggl(\frac{x}{z},M^2\biggr) \, 
\nonumber \\ && \qquad \times 
\biggl[ \frac{1+z^2}{1-z}
\biggl(\ln\frac{M^2}{m_q^2}-2\ln(1-z)-1\biggr) \biggr]_+\;,
\end{eqnarray}
where $q(x,M^2)$ can be taken directly from existing PDF
parameterizations in the $\MSbar$ scheme.
In the approach with subtraction from PDFs it is easy to keep the
completely differential form of the sub-process cross section and
therefore to impose any kind of experimental cuts.

The results of calculations were automatically transformed into software modules in the {\tt FORTRAN}
language with the standard {\SANC} interface.
Universal one-loop modules were created for scalar form factors and helicity amplitudes. 
Evaluation of loop integrals in the modules is performed by 
{\tt Looptools}~\cite{Hahn:1998yk} 
and {\tt SANClib}~\cite{Bardin:2009zz,Bardin:2009ix} packages.
The modules are independent and publicly available on the {\SANC} website. 

\underline{\bf Higher order QED corrections} 

The leading logarithmic approximation was applied to take into account
the QED corrections of the orders 
${\cal O}(\alpha^n L^n), n = 2, 3$~\cite{Antropov:2017bed}. 
Both photonic and light pair corrections were treated in this way.

\underline{\bf Systematic treatment of the photon-induced contribution}

DY processes allow accessing final states with very high di-lepton invariant masses.
It this case, the photon-induced sub-process contributions 
{$q \gamma \to q \ell^+ \ell^-$}, 
{$\gamma \gamma \to \ell^+ \ell^-$}
with photon-type partons being found inside protons 
become numerically relevant together with the standard quark--antiquark annihilation
sub-processes. Contributions to the NC DY cross-section due to the
photon-induced process $ \gamma \gamma \to \ell^+ \ell^- $ can reach up to
$10-20 \% $ for high invariant mass $ M_{\ell^+\ell^-} $ with a choice of
kinematic cuts typical of LHC experiments.
The first evidence of this kind of background was found by the ATLAS
Collaboration in high-mass NC DY cross section measurements~\cite{Aad:2013iua}.
Photon-induced Drell--Yan processes were carefully investigated in our paper~\cite{Arbuzov:2007kp}.

\underline{\bf Factorization scheme and scale dependence}

Factorization scheme and scale dependence is unavoidable in calculations of high-energy
hadronic processes. One can reduce it by adjustment of the scheme and scale choices 
and by including relevant higher order effects. The corresponding uncertainties were
estimated by comparison of numerical results in three EW schemes ($\alpha(0)$, 
$G_\mu$, and $\alpha(M_Z)$) for different factorization scales, see details in~\cite{Bondarenko:2013nu}. 

\underline{\bf Electroweak corrections beyond NLO approximation}

Theoretical uncertainties mainly come from the higher orders of perturbation theory contributions. 
Two-loop EW corrections, the EW Sudakov logs, or two-loop vacuum polarization and multiple final state radiation are of greatest importance. 
We have already implemented the leading
in $G_\mu m_t^2 $ two-loop EW and mixed EW$\otimes$QCD radiative corrections through the
$\Delta\rho$ and $\Delta\alpha$ parameters
\cite{Arbuzov:2015yja}.
Now we are implementing into \ReneSANCe and \MCSANC the
mixed QCD$\otimes$QED 
${\cal O}(\alpha_s \alpha)$ and pure QED ${\cal O}(\alpha^2 )$ corrections following ~\cite{deFlorian:2018wcj}. 
The corresponding study is in progress. 

\underline{\bf PDFs parameterization}
   
The present accuracy of PDFs in the kinematical region
relevant to the LHC leads to a huge uncertainty in DY cross section of the order of 5\% or even more.
The situation will be improved only after new fits of PDFs based on the LHC data.

\underline{\bf Matching procedure}

To get the required precision, we need an advanced implementation in Monte Carlo codes
matching procedure. Work is currently underway to implement 
the necessary matching procedure, i.e., precise corrections coordinated with
logarithmic (multiphoton) corrections of higher order.

\underline{\bf Polarization effects}

For the first time the study of spin effects in production of the massive vector boson
in collisions of longitudinally polarized hadrons at NLO EW level
for NC DY process is performed in our work~\cite{Bondarenko:2022zse}.
We estimate the polarization effects for single- $A_L$ and double-spin  asymmetries $A_{LL}$, 
cross sections and found significant contributions. Theoretical predictions on spin effects in
production of the massive vector boson
in collisions of longitudinally polarized hadrons are required for the polarized PDFs building. Such job can
be carried out by the {\tt xFitter}~\cite{Alekhin:2014irh}. In this case for polarized PDFs fitting they will use the data 
picked from RHIC collider which has longitudinal polarization mode of operation.
Here we show relative corrections $\delta A_{L}$, in $\%$ for the
$Z$ boson rapidity ${\mathrm y}_{\mathrm \sss Z}$ and the lepton pseudorapidities $\eta_{\ell^\pm}$ distributions. 
\begin{figure*}[!h]
\begin{tabular}{ccc}  
    \includegraphics[width=0.34\textwidth]{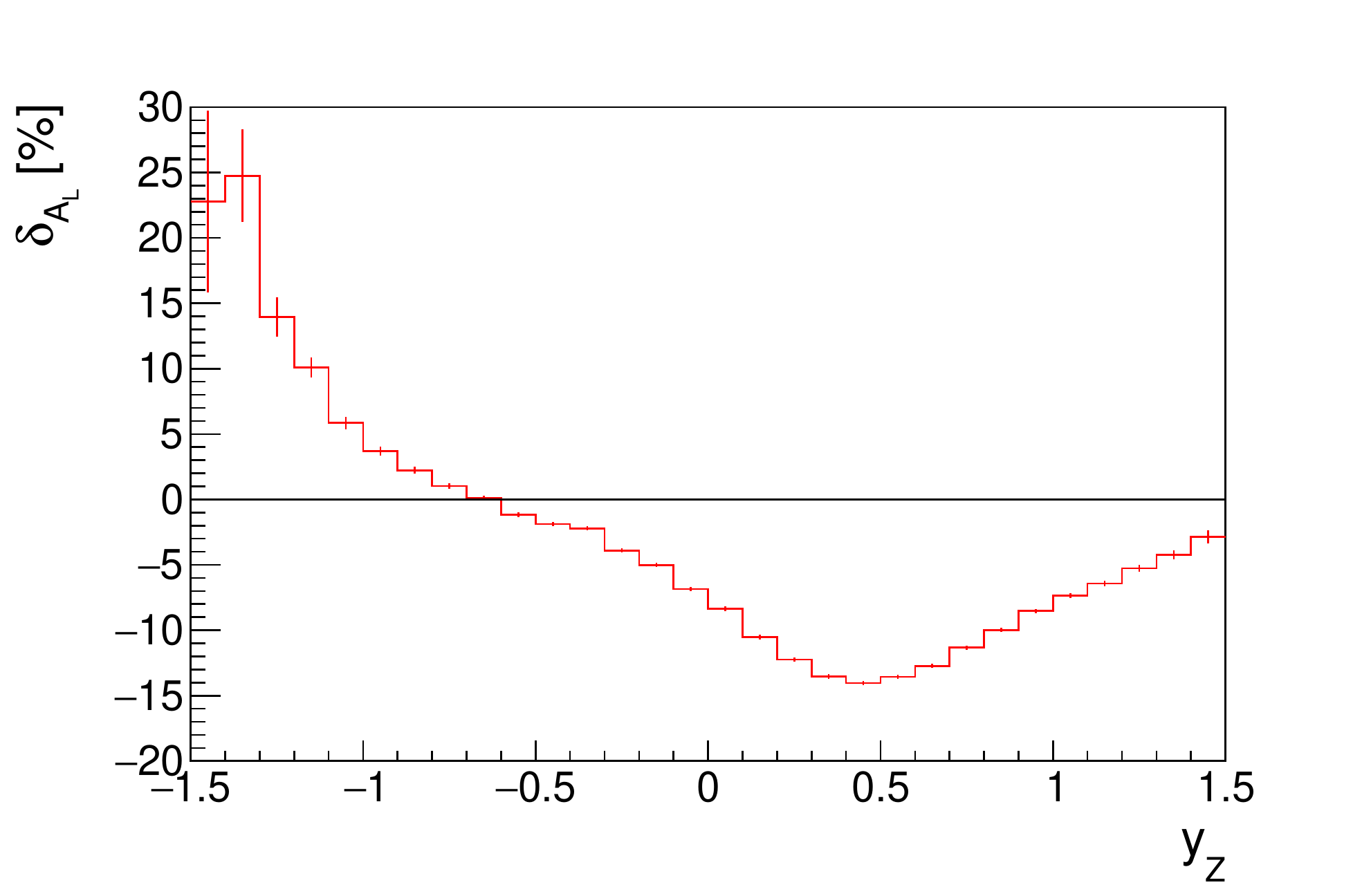}
    \hspace*{-3mm}
     \includegraphics[width=0.34\textwidth]{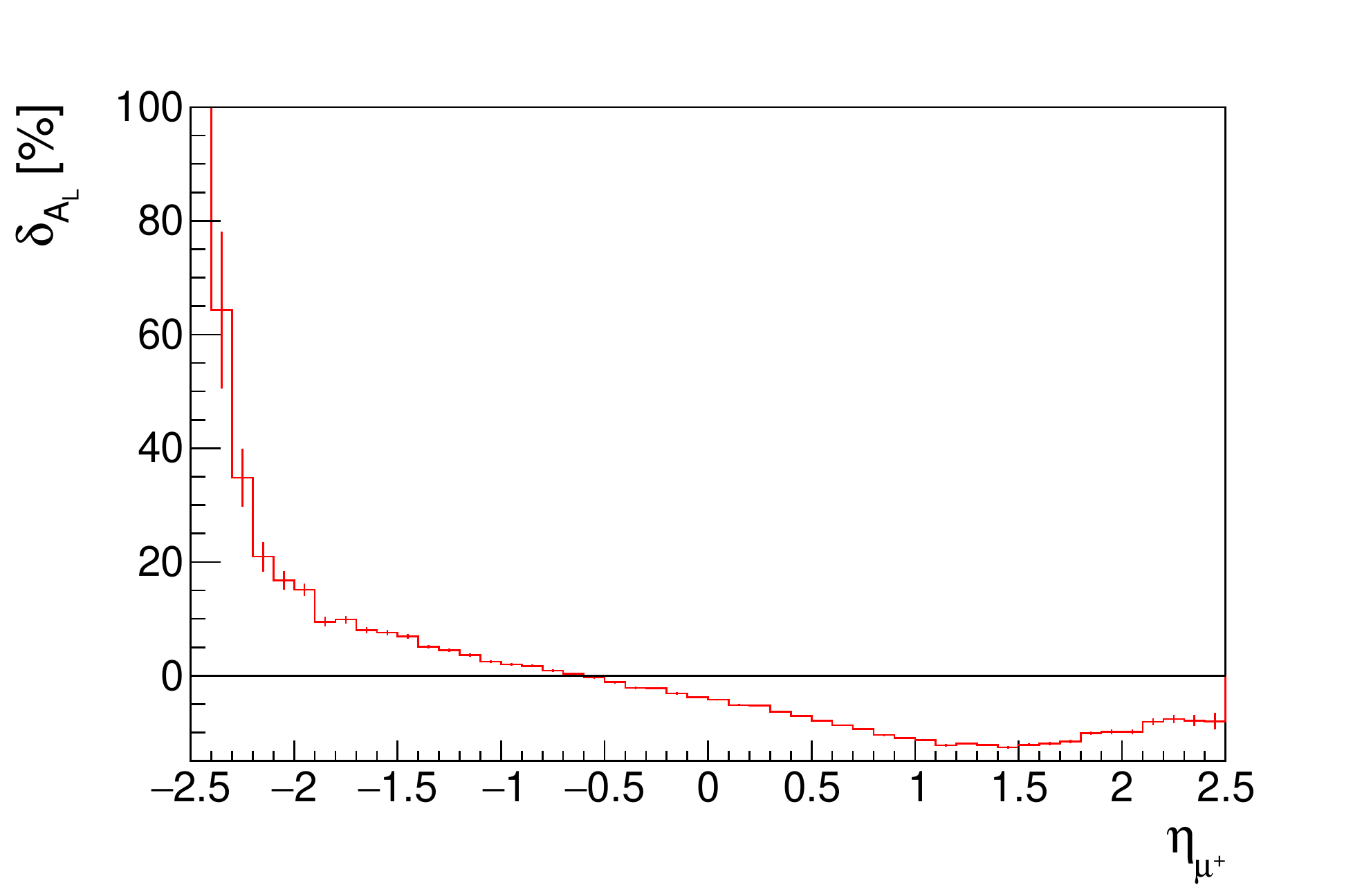} 
     \hspace*{-3mm}
    \includegraphics[width=0.34\textwidth]{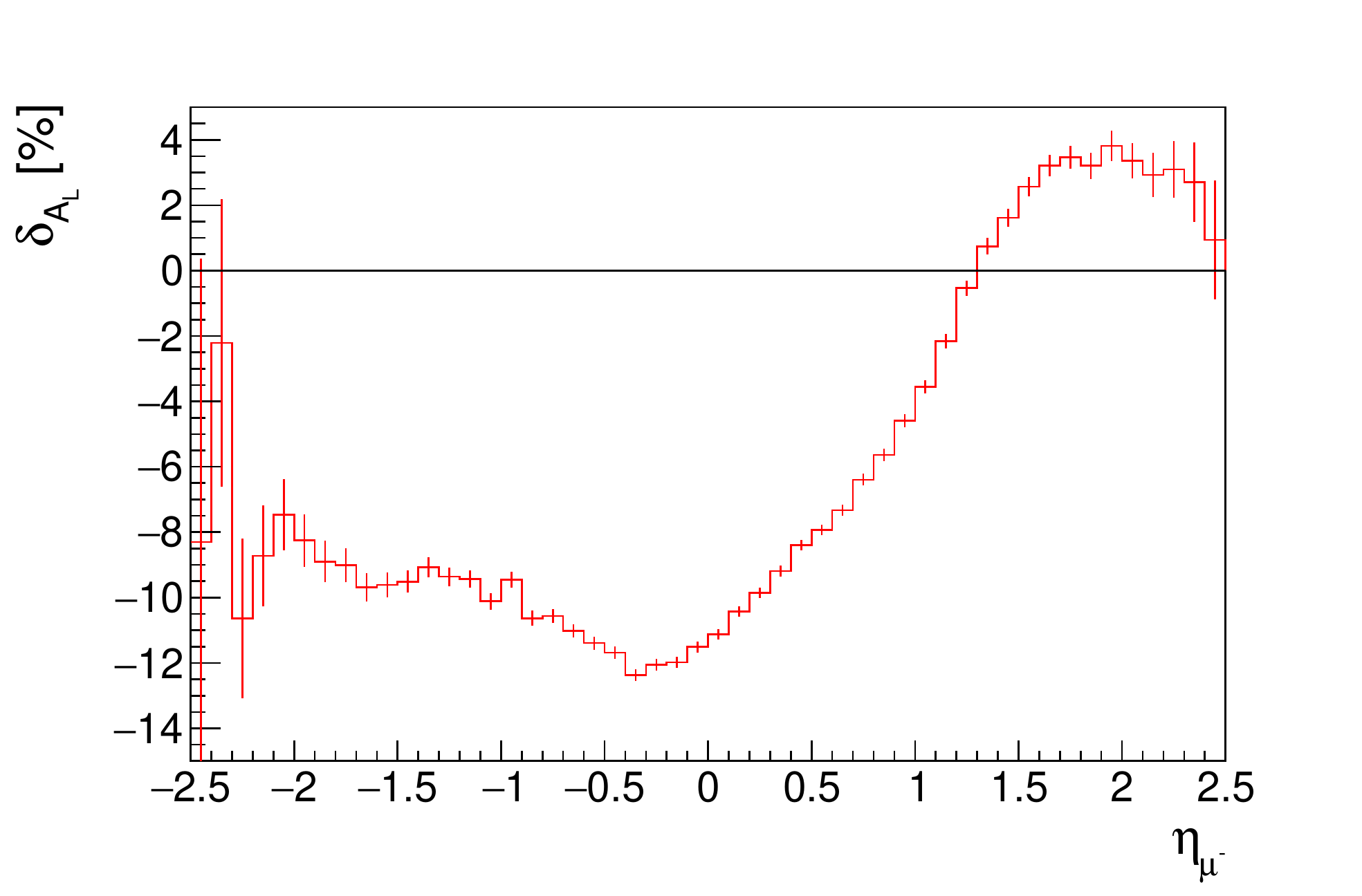}
\end{tabular}
\end{figure*}

\section{Conclusion}
The current versions of {\ReneSANCe} and {\MCSANC} are adjusted to studies of various effects due to EW radiative corrections to realistic observables at LHC, Tevatron and RHIC for massive vector boson production in neutral current channel.

\section*{Funding}
\label{sec:funding}
The research is supported by the Russian Science Foundation (project No. 22-12-00021). 

\bibliographystyle{pepan}
\bibliography{dimuons_SANC_hep}
\end{document}